\documentclass[fleqn,usenatbib]{mnras}

\usepackage[T1]{fontenc}

\usepackage[normalem]{ulem}

\DeclareRobustCommand{\VAN}[3]{#2}
\let\VANthebibliography\thebibliography
\def\thebibliography{\DeclareRobustCommand{\VAN}[3]{##3}\VANthebibliography}

\usepackage{graphicx}	
\usepackage{amsmath}	
\usepackage{amssymb}	

\usepackage{newtxtext,newtxmath}

\newcommand{\Msun}{{M_{\odot}}}
\newcommand{\concsection}[1]{ \vspace{3mm}
\noindent \textbf{\textit{{#1}}:}}

\title[Emulating tSZ images]{Emulating Sunyaev-Zeldovich Images of Galaxy Clusters using Auto-Encoders}

\author[T. Rothschild et al.]{
Tibor Rothschild,$^{1}$\thanks{E-mail: tibor.rothschild@yale.edu}
Daisuke Nagai,$^{1,2}$
Han Aung,$^{1}$
Sheridan B. Green,$^{1}$
Michelle Ntampaka,$^{3,4}$
John ZuHone$^{5}$
\\
$^{1}$Department of Physics, Yale University, New Haven, CT 06520, USA \\
$^{2}$Department of Astronomy, Yale University, New Haven, CT 06520, USA \\
$^{3}$Space Telescope Science Institute, Baltimore, MD 21218, USA\\
$^{4}$Department of Physics \& Astronomy, Johns Hopkins University, Baltimore, MD 21218, USA\\
$^{5}$Chandra X-Ray Center, 60 Garden Street, Cambridge, MA, 02138, USA\\
}

\pubyear{2021}

\begin{document}
\label{firstpage}
\pagerange{\pageref{firstpage}--\pageref{lastpage}}
\maketitle

\begin{abstract}

We develop a machine learning algorithm that generates high-resolution thermal Sunyaev-Zeldovich (SZ) maps of novel galaxy clusters given only halo mass and mass accretion rate. The algorithm uses a conditional variational autoencoder (CVAE) in the form of a convolutional neural network and is trained with SZ maps generated from the IllustrisTNG simulation. Our method can reproduce many of the details of galaxy clusters that analytical models usually lack, such as internal structure and aspherical distribution of gas created by mergers, while achieving the same computational feasibility, allowing us to generate mock SZ maps for over $10^5$ clusters in 30 seconds on a laptop. 
We show that the model is capable of generating novel clusters (i.e. not found in the training set) and that the model accurately reproduces the effects of mass and mass accretion rate on the SZ images, such as scatter, asymmetry, and concentration, in addition to modeling merging sub-clusters. This work demonstrates the viability of machine-learning--based methods for producing the number of realistic, high-resolution maps of galaxy clusters necessary to achieve statistical constraints from future SZ surveys. 
\end{abstract}




\section{Introduction}
\label{sec:intro}

As the largest gravitationally-bound objects in the universe, galaxy clusters are a key tool in the field of cosmology. The formation and evolution of such massive objects is a direct consequence of the abundance of dark matter and dark energy in the Universe \citep[see ][for reviews]{allenConstrainCosmo,kravtsovClusterFormation}. Concretely, one can constrain cosmological parameters by measuring the halo mass function. Thus, one of the key steps in constraining cosmology is to measure the halo mass, which is usually achieved through observed properties which scale with mass \citep[][for a recent review]{Pratt19}.

One such powerful method is to measure the integrated intensity deviations from the cosmic microwave background (CMB) due to the Sunyaev-Zeldovich (SZ) effects \citep[e.g.,][and reference therein]{carlstromSZ,planck15}. The hot intracluster medium (ICM) boosts CMB photons to higher frequencies resulting in a characteristic increase or decrease in CMB intensity depending on the specific microwave frequency being observed. The change in intensity does not suffer from cosmological dimming which is an advantage over other methods, such as X-ray observations, for capturing data at higher redshifts. The SZ effect generally sees its most dominant contribution from the thermal motion of ICM electrons so the magnitude of the thermal SZ effect is proportional to the line-of-sight integrated thermal pressure \citep{szOG}. The integrated thermal SZ effect over an area of sky where a cluster is located will be proportional to the cluster's mass. This relationship, termed the $Y_{SZ}-M$ scaling relation, is a valuable tool for estimating halo mass \citep{mroczkowskiSZ} from which one can readily determine cluster abundance. Its ability to detect high redshift clusters makes it a critical tool for capturing a longer and more accurate picture of structure growth in the universe, which provides useful information on the nature of dark energy.

However, the details of this scaling relation, such as its normalization and scatter (the inter-cluster variation in integrated $Y_{SZ}$ flux at fixed mass), are sensitive to the mass accretion rate of the cluster because the turbulence introduced by structure formation processes adds additional pressure support \citep{krauseMergerScatter,yuYMScatter,greenMARSZ}. This causes a hydrostatic mass bias in measuring mass from SZ surveys \citep{planck14_cluster_mass,Nelson14,shi2016_hse}. The mergers and recent accretions also tend to cause asphericity in the clusters which introduces additional scatter in projected measurements \citep{chenMAR,lau21_shape,Machado20_shape}. Baryonic physics, such as feedback, also adds turbulent pressure support, in addition to the structure formation processes \citep{battagliaYMRelation,pike2014}. Thus, one requires high-resolution cosmological hydrodynamical simulations to constrain the effects of both astrophysics and cosmological structure formation processes. 

The upcoming CMB-S4 \citep{cmbS4}, Simons Observatory \citep{soPaper}, and CMB-HD \citep{sehgal2019cmbhd} surveys will provide us with a wealth of new, high-resolution SZ maps of hundreds of thousands of clusters \citep{raghunathan21}. In order to maximize the scientific return of these observations, we need to simulate a similar number of clusters to achieve the required statistics while controlling the systematic uncertainties of baryonic physics and mass accretion history. N-body simulations are one relatively simple approach, but for obtaining predictions that exclude baryonic physics. 
The necessary complex hydrodynamical simulations for the additional task of modeling of baryonic physics requires sub-parsec resolution scales and simulation boxes of order Gpc. High-quality simulation data with detailed baryonic physics requires millions of CPU hours \citep[e.g.,][]{illustrisTNG2}, which ends up producing $>2$ orders of magnitudes fewer clusters than are expected to be observed in upcoming surveys \citep{soPaper}. This stands in contrast with the previously mentioned N-body simulations, which are much less resource-intensive to run but neglect baryonic physics.

Fortunately, there exist techniques for addressing this issue by using additional techniques on top of N-body simulations. In general, one would like to find a mapping between key halo characteristics, such as mass, and a corresponding thermal pressure profile. For example, the semi-analytic models of gas \citep[e.g.,][]{ks01SA,shaw10SA,Flender17} produce a thermal pressure profile given characteristics like halo mass. Previous work has used similar semi-analytic models to create all-sky light-cone maps for X-ray surveys such as eROSITA \citep{zandanel18}. Even in the absence of N-body simulations, one can use semi-analytic models to make predictions of observables using a halo mass function (HMF) in a halo-based approach \citep{multiWavelengthShirasaki}. Using a HMF, one can create a halo catalog containing appropriate sample of cluster mass and redshift pairs. This halo catalog can then be fed into a model which can generate maps of observables for each individual halo. A halo-based approach is particularly desirable because halo catalogs are in general much less computationally expensive than N-body simulations. The relatively small computational cost allows one to easily generate SZ maps for a large number of different halo catalogs corresponding to different cosmologies.

However, most semi-analytical models which would be used in a halo-based approach assume spherical symmetry and do not model the scatter introduced by secondary parameters of halos, such as mass accretion rate. Thus, by virtue of their simplifying assumptions, semi-analytic models still suffer from less predictive power. Concretely, they generally produce spherically symmetric distributions of pressure which for individual clusters is not necessarily the norm. It would be ideal if one could combine the detailed predictions of hydrodynamical simulations with the efficiency and minimal necessary inputs (i.e., only a few key parameters such as mass, MAR, and redshift) of semi-analytical models such that these detailed predictions could be used in a halo-based approach to the prediction of observables.

Machine learning (ML), and deep learning (DL) in particular, has recently emerged as a promising tool for modeling complex relationships in data for a variety of purposes \citep{dlNature}, including X-ray and SZ cluster cosmology \citep{Ntampaka19,Armitage19,Cohn_Battaglia20}. A growing body of research has been focused specifically on using astrophysical and cosmological simulation data to train machine learning models in an effort to teach them to \textit{create} data with simulation-level fidelity in negligible time \cite[e.g.,][]{ravanbakhshGenGal}. Several generative models have successfully reproduced images with a focus on low-resolution ($\geq 0.1$ Mpc), large scale maps, where the success is determined by reproducing power spectra \citep{paintingBaryons, teachingNeuralNetworks, heOverdensityToHalos, bernardiniOverdensityToHalos}. These approaches are not halo-based, and so using them requires matter density maps from N-body simulations rather than a few key halo characteristics. In this work, we instead seek to establish that deep learning can produce higher resolution ($\sim 30$ kpc) SZ signals around galaxy clusters, and assess it by searching for evidence of more subtle physics like mergers, accretion, and morphological asymmetry. Reproducing cluster-scale physics at this resolution would open up new opportunities for advancing our understanding of cosmology with the corresponding high-resolution surveys coming online in the near future. We also seek to establish a greater degree of flexibility in data generation, in that we would like to explicitly only create objects with specific choices of physical properties. This allows for, among other things, a greater degree of interpretability of the model. 

In this study, we develop a machine learning model to efficiently generate SZ images of individual galaxy clusters given only salient halo properties of mass and MAR of the galaxy cluster. We obtain the training dataset from the IllustrisTNG simulations, with details provided in Sec.~\ref{sec:data}. In Sec.~\ref{sec:methods}, we describe the conditional variational autoencoder, a neural network architecture introduced by \cite{cvaeOriginal} which was a modification of the variational autoencoder from \cite{vaeOriginal}, as well as how we assess the success of the model. In Sec.~\ref{sec:results}, we present the results of our evaluation of the trained model. We interpret our results and discuss limitations in Sec.~\ref{sec:discussion}. Finally, in Sec.~\ref{sec:conclusions}, we conclude this work with a review of its results and implications.

\section{Data}
\label{sec:data}
In this section, we describe how we generate high resolution SZ maps based on the IllustrisTNG hydrodynamical cosmological simulations and the observables we derive from the SZ map.

\subsection{Simulation}
The IllustrisTNG simulations are a suite of magnetohydrodynamical simulations that incorporate complex baryonic processes. The simulations use a moving-mesh code AREPO \citep{arepo} to solve the equations of gravity and magneto-hydrodynamics. TNG simulations model the radio mode black hole feedback physics, star formation, supernova feedback, and chemical evolution, as well as ideal MHD. For more details on the simulations, we refer the readers to \citet{illustrisTNG1, illustrisTNG2, illustrisTNG3, illustrisTNG4, illustrisTNG5}. The simulations adopt a cosmology consistent with the latest constraints from \citet{Planck15_cosmo}, where $\Omega_m=0.3089$, $\Omega_{\Lambda}=0.6911$, $h=0.6774$, $\Omega_b=0.0486$, $\sigma_8=0.8159$, $n_s=0.9667$. 

The IllustrisTNG simulation suite consists of three primary simulations of different box sizes. We will focus on the TNG300-1 simulation which offers us the highest number of galaxy clusters at the highest resolution. It has a box size of $205$ Mpc/h. Dark matter and baryonic particles have mass resolutions of $5.9\times 10^7\Msun$ and $1.1\times 10^7\Msun$, respectively. Collisionless particles (dark matter and stars) have a softening length of $1.5$ kpc whereas gas particles have variable softening with a minimum of $370$ parsecs.

\subsection{Halo Catalog}
The halos in IllustrisTNG 300-1 are identified with the Subfind algorithm \citep{subfind}. The center of the FoF halos is the same as the center of the most massive subhalo within the FoF group. We select 296 halos where the most massive subhalo has a mass above $10^{14} M_{\sun} / h$. The spherical overdensity radii and masses such as $M_{200c}$ are then calculated using all of the halo particles.

The halos are then linked together through different snapshots with the Sublink algorithm, which tracks dark matter, stellar, and star forming gas particles \citep{sublink}. We define mass accretion rate (MAR) as $\Gamma \equiv \Delta \log(M_{200c})/\Delta \log(a)$ computed over the last dynamical time, $t_{\rm dyn} = 2\sqrt{R_{200c}^3/GM_{200c}}$ \citep{sparta}.

\subsection{SZ Map Generation}
At the $z=0$ snapshot, we compute 4 Mpc-wide maps of the change in intensity due to the thermal SZ (tSZ) effect with relativistic corrections for 3 orthogonal projections along the $x,y,$ and $z$ axes of each halo. We use yt \citep{ytProject} to perform the projections and SZpack \citep{szPack} to compute the tSZ signal, which is proportional to the Compton $y$ parameter
\begin{equation}
    y = \frac{\sigma_T}{m_ec^2}\int n_ek_B T_e  {\rm d} l ,
\end{equation}
where $n_e$ is the electron number density and $T_e$ is the electron temperature. To make these projections for a given cluster in our sample, we first count all the gas particles within the FoF group of the halo. We then make a cut on gas temperature, only including particles with $T > 10^{5.3}$~K, since it is only the particles representing hot and ionized gas that we expect to make a contribution to the tSZ signal. For these particles, the gas is fully ionized and we simply set $n_e = \rho/(\mu_e{m_p})$, with $\mu_e = 1.14$, the mean molecular weight for electrons. Following \citet[][their Section 3.2]{Chluba2013}, we make projections of the Compton optical depth $\tau$, the $\tau$-weighted mean $T_e$, and the $\tau$-weighted variance of $T_e$, which are used in the construction of the tSZ signal by SZpack. The projections of these quantities are made by assigning values to pixels in the map using the smoothing length of the gas particles and the standard SPH cubic smoothing kernel.

SZpack also computes the relativistic corrections to the tSZ signal, which we include in our maps. We do not include the kinematic SZ effect. In this work we focus on the tSZ effect, and so choose to make `observations' at 90 GHz where the tSZ effect has the dominant contribution and the corrections are minor at this relatively low frequency. This yields a dataset of $3 \times 296 = 888$ SZ maps that we then split $90-10$ into train and validation sets. The final maps have a $128 \times 128$ resolution in pixels, meaning each pixel is $\approx 30$ kpc, ensuring that the pixels at the cluster outskirts contain enough gas particles for a smooth map. For reference, this corresponds to an angular resolution of $\approx 16$ arcsec at $z=0.1$. We also train on horizontally and vertically flipped maps, as well as rotations by multiples of $90 \degr$. Therefore we use an effective dataset with $888 \times 8 = 7104$ maps.

The unprocessed SZ maps have an extremely high dynamic range, spanning 18 orders of magnitude. If we train on this scale, we find that the network unsurprisingly learns to neglect the smaller values occurring on the outskirts of the clusters. Therefore we perform the log transform 
\begin{equation}
a' = \log_{\frac{1}{\epsilon}}\left(\frac{a + \epsilon}{\epsilon}\right)
\end{equation}
where $a$ is the unprocessed pixel value and $\epsilon$ is a small positive constant, both of which are in units of MJy sr$^{-1}$. For this work, we chose $\epsilon = 10^{-6}$ because $90\%$ of the pixel values in the dataset are larger than this value, so it was determined to be a good balance of retaining data and preserving contrast.

Masses ($M$) in units of $M_{\sun} h^{-1}$ are input to the model only after undergoing the transform
\begin{equation}
M' = 10^{\log_{10} (M) - 13}.
\end{equation}
However, we will not quote masses in this form in this work.

\subsection{Computing Observables}
\label{sec:obs}
The key observable quantity derived from an SZ map is the integrated Compton-y which can be expressed as:
\begin{equation}
    Y=\int y {\rm d}A \propto \int n_e T {\rm d}V.
\end{equation}
Therefore $Y$ is in fact proportional to $M_{\rm gas}T$, which is a robust proxy of halo mass \citep{kravtsov_etal2006,nagai2006}. We compute this quantity from the images as the sum of all pixel values within the projected radius of $R_{200c}$, a quantity directly proportional to $Y$ modulo relativistic corrections. $R_{200c}$ was chosen for the mass and MAR definitions because it was largest radius that was feasible given the field of view of the computed SZ maps used in this work. It is also frequently used as a radius and mass definition in observational SZ science \citep{actSZ}.

The mass accretion rate of the clusters dictates the morphology of the cluster and thus, the SZ images. Earlier forming clusters have more mass concentrated in the center of the cluster \citep{Ludlow2016}, while the late forming clusters are usually more aspherical due to mergers \citep{chenMAR}. Therefore, to check whether our algorithm can reproduce the effects of mass accretion rate on the SZ images, we examine the asymmetry and concentration of the SZ images, which have been used to quantify the morphology of X-ray cluster images \citep{lotz04,Rasia13,Lovisari17}. These can be calculated directly from the SZ map, a 2-dimensional matrix denoted by $K$, and the halo mass (or radius). For consistency with previous work, we use the morphological parameter definitions from \cite{greenMorph} where pixels within $R_{500c}$ are used to compute parameter values.

Asymmetry: $A = \frac{|K-K_{180}|}{|K|}$ where $|K|$ is the sum of the absolute values of the entries of $K$ and $K_{180}$ is the $180 \degr$ rotation of $K$.

Concentration: $C = \frac{|K_{0.1*R200c}|}{|K|}$ where a subscript of $k$ denotes setting all values outside the circle of radius $k$ to 0.

We also define an additional parameter, smoothness: $S = \frac{|K-K_{S}|}{|K|}$ where $K_S$ denotes $K$ after undergoing boxcar smoothing with a width of $0.25R_{500c}$ which results in an average kernel width of about 6 pixels. This parameter is useful for two reasons: it quantifies the degree of small-scale substructure within the cluster, the presence of which is positively correlated with a violent mass accretion history, and it tells us whether the algorithm can produce maps of same resolution as the training data.

\section{Methods}
\label{sec:methods}
\subsection{Conditional Variational Autoencoder}
Machine learning techniques are generally used to create representations of salient structure in a given training dataset, and then often utilize these representations to perform some task. In generative modeling, the goal is to use the learned representation to create data that in some sense resembles the original training data. Convolutional neural networks have excelled in learning representations of image data, and so they have been successfully used together with generative modeling approaches when working with image data \citep{ganOriginal, vaeOriginal}.

One technique used for generative modeling is the conditional variational autoencoder (CVAE), introduced in \cite{cvaeOriginal}. A variational autoencoder consists of an encoder and a decoder. The encoder is trained to learn to represent (encode) the image data as data points in a distribution in a relatively low dimensional latent parameter space. After training, it can then sample randomly from a Gaussian distribution in this lower dimensional space and the decoder transforms (decodes) these parameters back into an image. A CVAE allows one to control some of the latent parameter space and use the known properties of the image as an input feature when generating an image, while randomly sampling the rest of the parameter space. This provides us with an ideal algorithm to produce a cluster image with a desired physical property such as mass or mass accretion rate. 

\begin{figure*}
	\includegraphics[width=\textwidth]{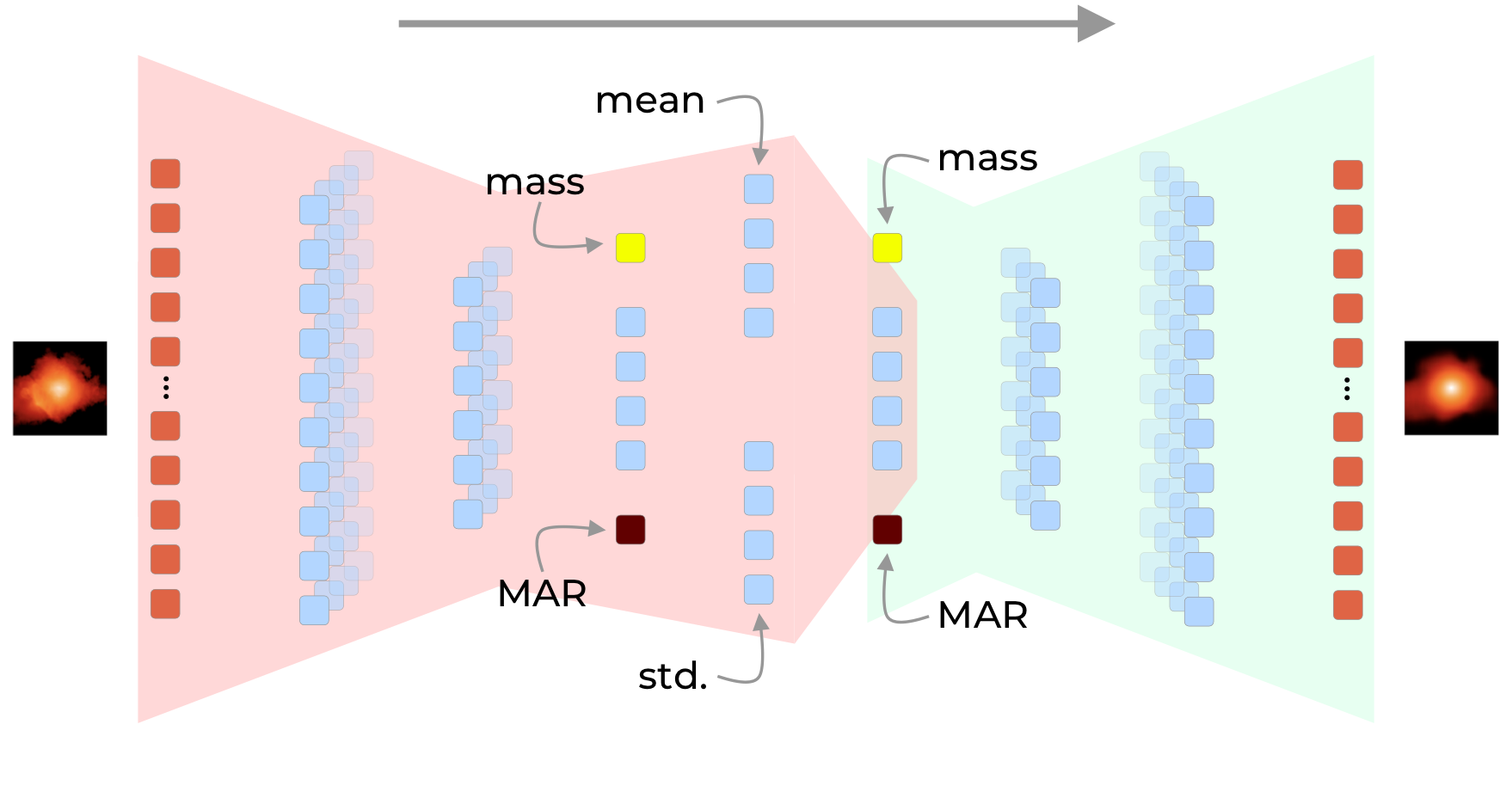}
    \caption{A visualization of our CVAE architecture. The red and green backgrounds signify the encoder and decoder, respectively. The arrow shows the direction of data flow through the network. During training time, the network is given an SZ map, shown on the far left, and must compress it before producing another SZ map, shown on the far right, ideally looking as close as possible to the input. During test/evaluation time, a sample from a Gaussian (along with the desired mass and MAR) is used as input to the decoder and the encoder is unused.}
    \label{fig:architecture}
\end{figure*}

\begin{table}
	\centering
	\caption{Each row represents a single layer of the network. They are ordered in the order that they process inputs, with the exception of the starred pair of rows which receive identical data from the last unstarred row and combine their respective outputs in the next unstarred row via the reparametrization trick \citep{vaeOriginal}. The conv. column specifies whether the layer is convolutional or not and the f.c. column indicates whether or not it is a standard fully connected layer. A checkmark in the concat. column signifies that the inputs to the layer are concatenated with the mass and MAR.}
	\label{tab:architecture}
	\begin{tabular}{lcccr} 
		\hline
		\# of layers & \# of filters or \# of neurons & conv. & f.c. & concat.\\
		\hline
		$2 \times$ & 64 & \checkmark \\
		$2 \times$ & 128 & \checkmark\\
		$3 \times$ & 256 & \checkmark\\
        $1 \times$ & 256 & & \checkmark & \checkmark \\
        $1 \times$ & 256 & & \checkmark \\
        $(\ast) ~ 1 \times$ & 16 & & \checkmark \\
        $(\ast) ~ 1 \times$ & 16 & & \checkmark \\
        $1 \times$ & 16 & & \checkmark \\
        $1 \times$ & 18 & & \checkmark & \checkmark \\
        $1 \times$ & 256 & & \checkmark \\
        $3 \times$ & 256 & \checkmark\\
        $2 \times$ & 128 & \checkmark \\
		$2 \times$ & 64 & \checkmark\\
        $1 \times$ & 1 & \checkmark\\
		\hline
	\end{tabular}
\end{table}

\subsection{Model Architecture}
In this work, we utilize the general form of the original CVAE architecture, with a few modifications. An approximate visual representation is shown in Fig.~\ref{fig:architecture}, and the further particularities of our configuration are given in Table~\ref{tab:architecture}. Every convolutional layer in the encoder uses $2 \times 2$ strides and every convolutional layer in the decoder is followed by $2 \times 2$ upsampling (except for the final one). All convolutional layers use $3 \times 3$ filter sizes. We use ReLU as the activation function for all CNN layers except the final layer of the decoder, which uses the absolute value function. We found this to cause training converge much faster and avoid poor local optima, as opposed to no activation function or common alternatives like $\tanh$ or sigmoid. 

Additionally, during training we feed the network the mass and MAR of the clusters at the end of the encoding stage and at the beginning of the decoding stage in an effort to force it to utilize this information to create more optimal encodings and recreate the training cluster. Then during test time, one can sample randomly from the learned distribution \textit{and} feed in a specific mass and MAR, such that ideally the network will then "decode" these inputs into a new cluster with the specified mass and MAR.

In order to fix hyperparameter values, we generally proceed by testing a range of values and then progressively refine this range by identifying trends in performance with respect to the hyperparameter in question. For example, a key parameter of the model is the size of the final compressed representation of the input data. Some intuition is that a smaller size will push the model towards underfitting, while a larger size can result in a kind of overfitting. We explored sizes ranging from 2 neurons to 256 neurons, and ultimately settled on 16 based on the its superior ability to minimize the loss function on the validation set.

The model's loss function is standard for the (C)VAE architecture: the sum of a mean squared error pixel-wise reconstruction loss and Kullback-Leibler (KL) divergence loss \citep{cvaeOriginal}. The KL divergence term measures the degree to which the encodings produced by the model "look like" a unit Gaussian. The inclusion of the term encourages the model to produce encodings which could have been sampled from a unit Gaussian. The model is trained using the Adam optimizer \citep{adam} with an initial learning rate of $10^{-4}$ and batch size of $8$. We additionally decay the step size parameter $\alpha$ used by the Adam optimizer by $0.2$ when the validation loss fails to improve for $40$ epochs. Training is stopped when the validation loss fails to improve for $80$ epochs.

\subsection{Evaluation}
\label{sec:evaluation}
In order to evaluate the success of the model, we will use the physical quantities introduced in section \ref{sec:obs} to test how ``realistic'' the images produced by the model really are. It should be as difficult as possible to distinguish a cluster generated by the model from a simulated cluster. But we do not necessarily mean ``distinguish'' in a strictly visual sense. Rather, the generated clusters should have the same kinds of physical features as the simulated clusters. When training, this goal is encouraged by minimizing the pixel-wise loss, but as a general matter pixel-wise loss does not necessarily encourage the learning of physically-motivated features \citep{evalGenModels}. For this reason, we employ explicitly physically-motivated metrics in Sec.~\ref{sec:results} to evaluate the model. These metrics include the $Y_{SZ}-M$ relation, $Y_{SZ}-M$ residual--MAR relation, the MAR--concentration relation, and the distribution of morphologies, among others. In this section, we outline how exactly these metrics are calculated.

\subsubsection{Generating an equivalent population}
The model is successful in generating the novel clusters if randomly generated clusters from the model reproduce the same mass and mass accretion dependence as the original simulated clusters.
We are interested in comparing certain metrics aggregated over many clusters which we refer to as comparing two cluster ``populations'' (one population is taken from the training set, one is generated by the model). We would generally like the populations to be broadly equivalent i.e., have the same halo mass and MAR functions. From the set of all masses and MARs in our dataset, we randomly sample $N = 888$ of each and add noise to avoid the possibility of evaluating the model on the exact masses and MARs that it saw during training - essentially, deterring over-fitting. The noise for the two parameters $\log_{10}[M/(M_{\sun}/h)]$ and $\Gamma$ is sampled from a Gaussian distribution with $\mu=0$ and $\sigma=0.05$. This particular value of $\sigma$ was chosen because it is roughly two orders of magnitude smaller than the parameter values themselves. This means $\sigma$ is large enough to thwart any memorization by the model, but small enough to preserve the overall shape of the parameter distributions. We use these new perturbed masses and MARs as inputs to our model (along with Gaussian noise to serve as the latent representations). The model's output then serves as the generated counterpart ``population'' to the simulated ``population,'' which should have nearly the same halo mass and MAR function. Even when the simulated population is taken from the validation set, the fact that a different view of a cluster from the training set may appear in the validation set means that masses and MARs (not the image itself) from the training set may also be present in the validation set.

\subsubsection{Determining novelty}
In addition to ensuring the generated sample has the same physical dependence as the original sample, we also make sure the generated clusters are ``new'' and not a copy of the original clusters. If this aspect were neglected, the model could simply duplicate clusters from the training set in order to meet the realism goal, but this would be of no practical use. In some sense, the two criteria of realism and novelty are a reframing of the classic bias-variance trade-off in machine learning \citep{biasVariance}. In order to determine if a given cluster generated by the model is novel, we first normalize the generated image and every image in the training set such that each have unity as their maximum pixel value. Then we compute, using the Euclidean metric in pixel space, the 3 nearest neighbors in the training set to the generated cluster of interest. Afterwards, we perform a visual inspection to check if the generated cluster is in fact some kind of duplicate of one of the 3 nearest neighbors, a sign of over-fitting, and a potential failure mode.

\section{Results}
\label{sec:results}
\begin{figure*}
	\includegraphics[width=0.98\textwidth]{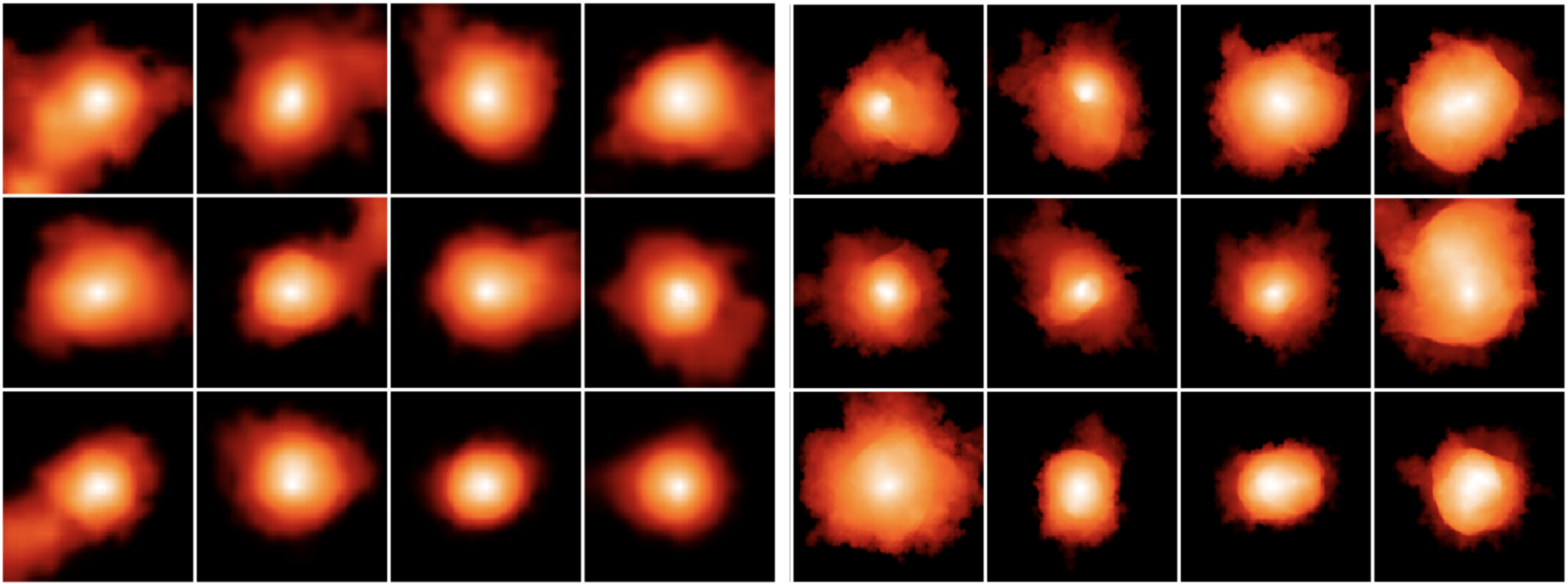}
    \caption{Sample of generated and simulated images with a width of 4 Mpc. \textit{Left panel:} cluster images generated by the model with a mass of $10^{14} M_{\sun}$ and a MAR of $4$. \textit{Right panel:} cluster images from simulation with masses and MARs closest to the previously referenced values of $10^{14} M_{\sun}$ and $4$. All clusters are shown on a log scale as described in Sec.~\ref{sec:data}.}
    \label{fig:examples}
\end{figure*}

\subsection{Reproducing the $Y_{SZ}-M$ relation}

While $Y_{SZ}$ is a robust proxy for the mass of the SZ clusters, the normalization and scatter of the $Y_{SZ}-M$ relation are strongly affected by the baryonic physics and structure formation processes. Mergers and mass accretion as well as feedback at the center of clusters generate non-thermal pressure which shapes the ways in which clusters deviate from the $Y_{SZ}-M$ relation. Therefore, capturing these deviations in the $Y_{SZ}-M$ relation of clusters generated by our model is a strong signal that the model is learning to effectively reproduce complex baryonic physics and mergers in simulated datasets. 

\begin{figure}
	\includegraphics[width=\columnwidth]{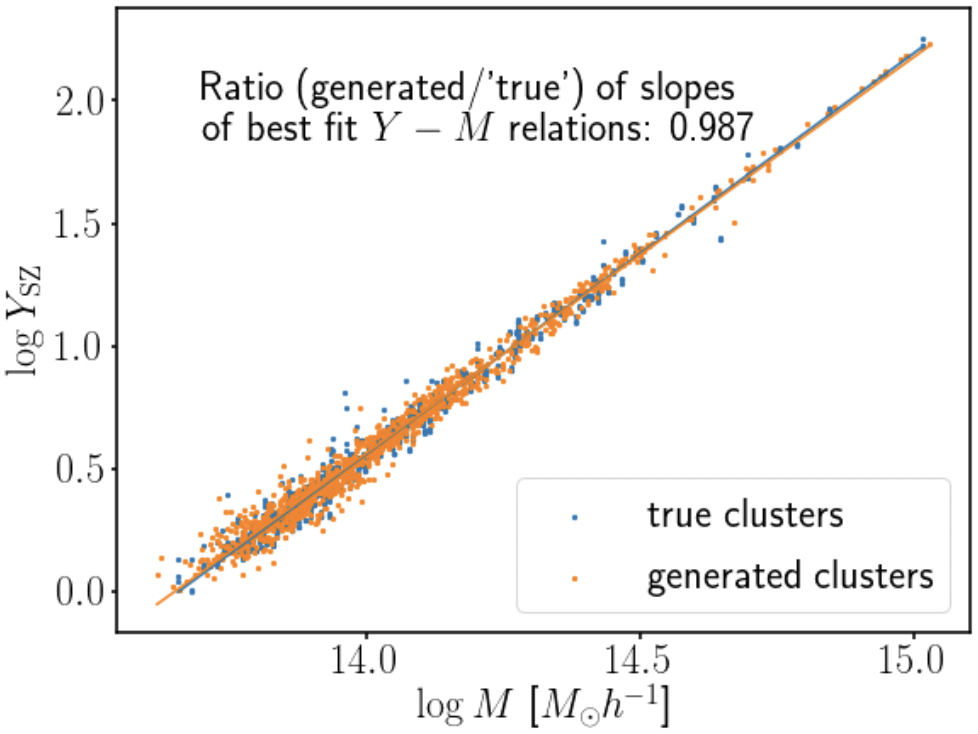}
    \caption{A comparison of the best-fit $Y_{SZ}-M$ relations for the two cluster populations. There is a strong linear correlation in the simulated data between $Y_{SZ}$ and $M$ due to the power law connecting the two quantities. We find that the model produces data which also has has this strong correlation. The ratio of slopes is one measure of the similarity in the correlations. 
    }
    \label{fig:Y-M_norm}
\end{figure}

\begin{figure}
	\includegraphics[width=\columnwidth]{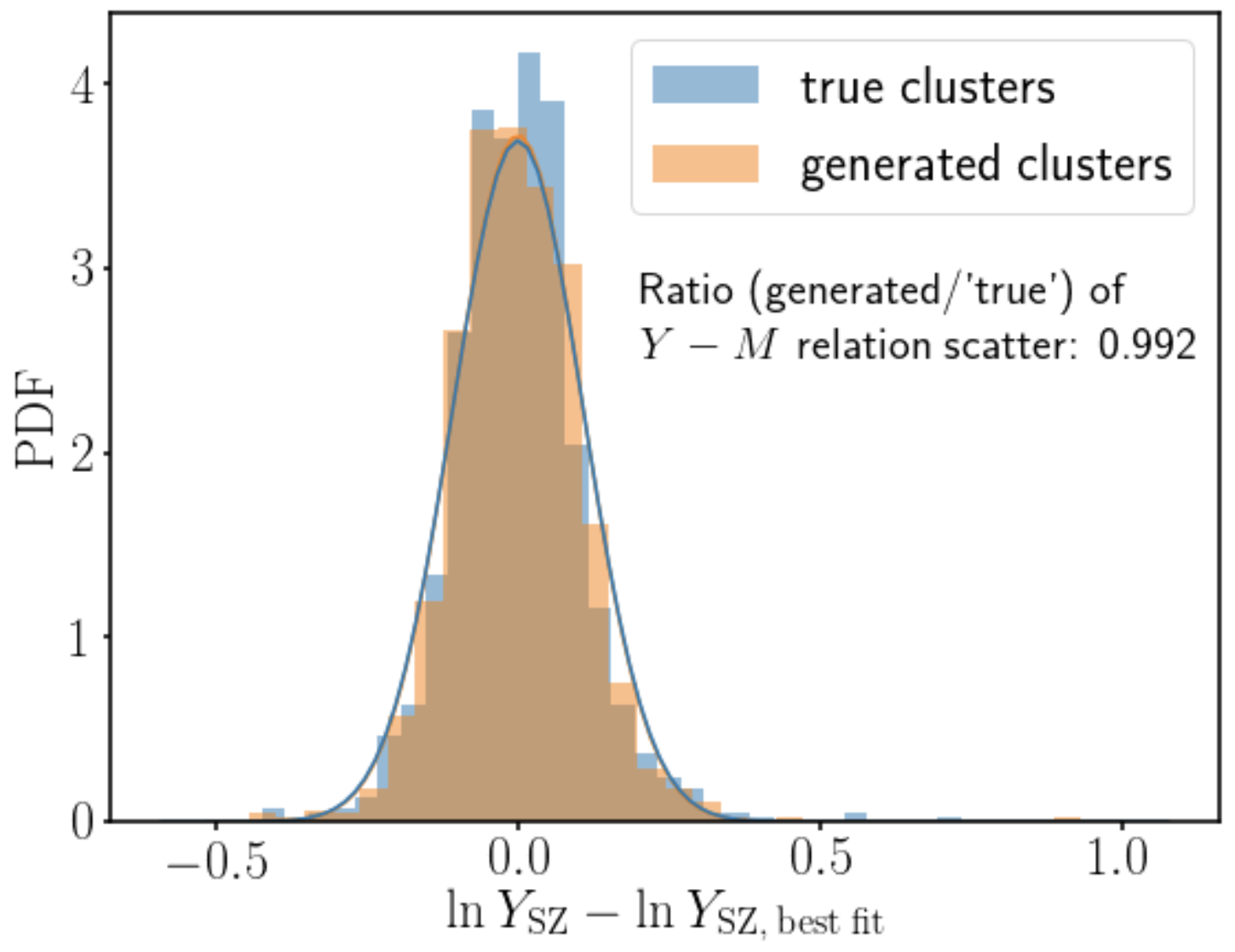}
    \caption{A comparison of the distribution of residuals of each cluster population from their own best-fit $Y_{SZ}-M$ relation, which are shown in Fig.~\ref{fig:Y-M_norm}. The plotted curves are the best-fit normal distributions for each population's residual distribution, from which the scatter is determined as the standard deviation. The scatter ratio measures similarity in the two populations' deviations from their $Y_{SZ}-M$ relation.}
    \label{fig:Y-M_scatter}
\end{figure}

We obtain a slope of $1.63$ and scatter of $10.8\%$ for $Y_{SZ}-M$ relation which is consistent with values found in previous studies of similar simulation data \citep{battagliaYMRelation,pike2014}. When comparing the normalization and scatter of the simulation data's $Y_{SZ}-M$ relation with that of a sample from the model, we find remarkable agreement. As shown in Fig.~\ref{fig:Y-M_norm} and Fig.~\ref{fig:Y-M_scatter}, the relations of the respective cluster populations have slope and scatter within $~1\%$ of each other. For context, previous semi-analytic models which only take into account mass accretion but not baryonic physics and clumpy accretions like major mergers \citep{greenMARSZ} reproduced $70\%$ of the intrinsic scatter when compared to non-radiative simulations, and reproduced just $1/3$ to $1/2$ of the scatter when compared to full-physics simulations.

\begin{figure}
	\includegraphics[width=\columnwidth]{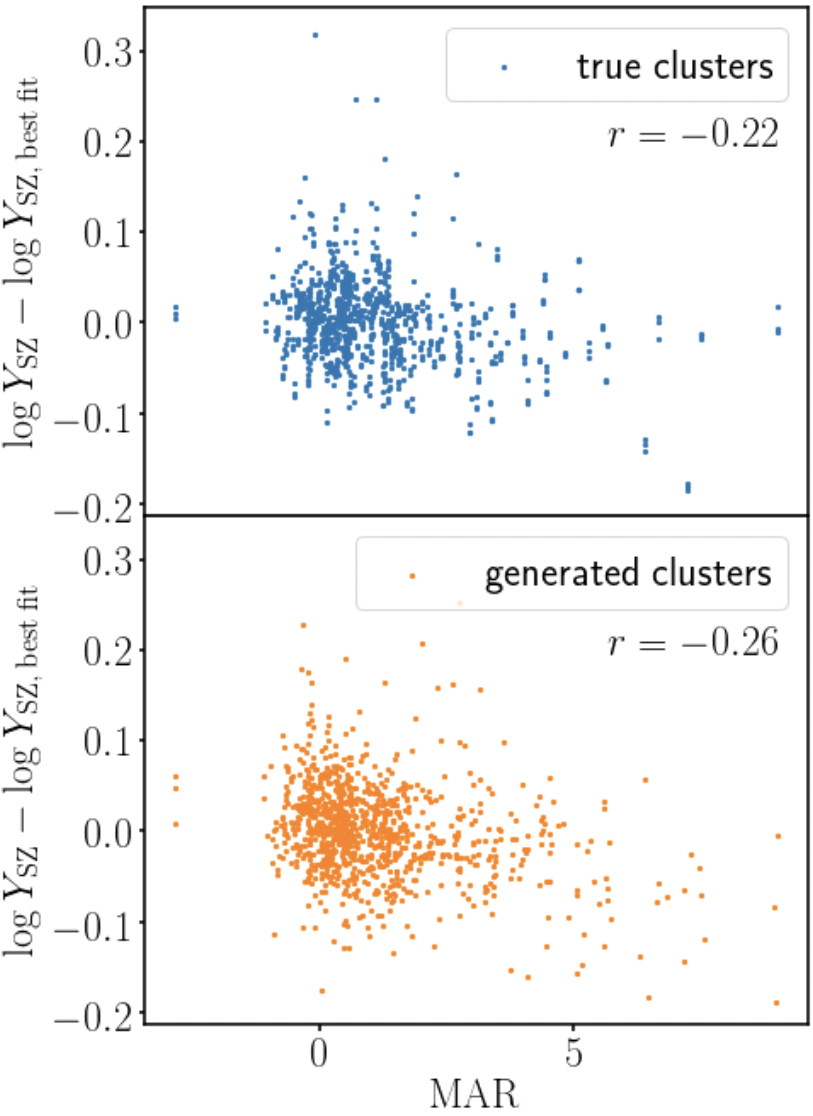}
    \caption{A comparison of the MAR - $Y_{SZ}-M$ residual relations for the simulated clusters above in blue and the clusters generated by the model below in orange. $r$ in each plot denotes the Spearman correlation coefficient between the MAR and $Y_{SZ}-M$ residual. There is a clear negative correlation for both populations of clusters. 
    }
    \label{fig:MAR-residual}
\end{figure}

By examining the effect of MAR on the $Y_{SZ}-M$ relation, we present evidence that our algorithm can reproduce not only the scatter of the population as a whole but also the source of scatter in individual clusters. Mergers and mass accretion introduce non-thermal pressure support, which reduces the SZ signal detected. Thus, at fixed mass, the higher the mass accretion rate, the higher the non-thermal pressure fraction and the smaller the $Y_{SZ}$ value \citep{nelsonNonThermal,greenMARSZ}. Thus, in general, we expect a negative correlation between the residual of the $Y_{SZ}-M$ relation and the mass accretion rate. We indeed find that the higher the residual, the smaller the mass accretion rate in the simulated SZ images as shown in the top panel of Fig.~\ref{fig:MAR-residual}, where the Spearman correlation coefficient is $r=-0.22$. Remarkably, we also find a similar correlation coefficient $r=-0.25$ between the residual and mass accretion rate of the ML generated clusters. This is a signal that the model explicitly emulates the effects of non-trivial physics associated with the mass accretion on SZ signal of the clusters. 

\subsection{Reproducing morphology}
In this section, we will explore how well the model reproduces the morphology of the SZ images of the clusters.

\begin{figure}
	\includegraphics[width=\columnwidth]{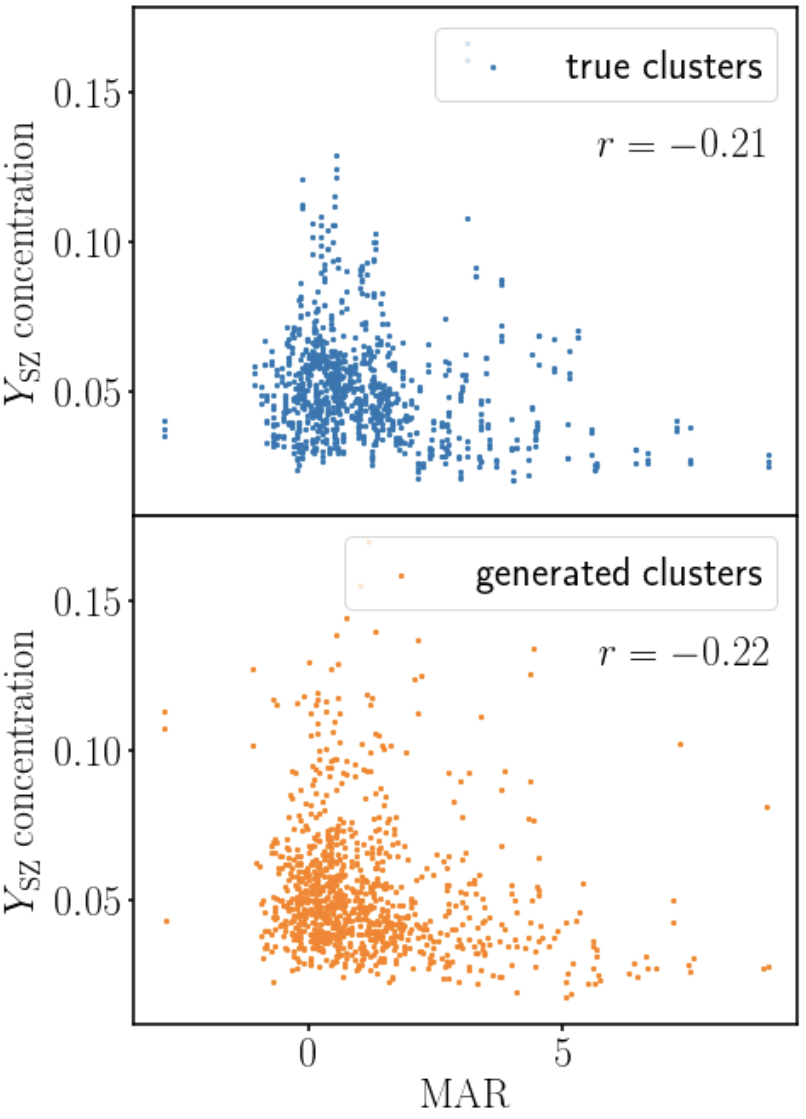}
    \caption{A comparison of the MAR-$Y_{SZ}$ concentration relations for the simulated clusters above in blue and the clusters generated by the model below in orange. $r$ in each plot denotes the Spearman correlation coefficient between the MAR and $Y_{SZ}$ concentration. There is a clear negative correlation for both populations of clusters. 
    }
    \label{fig:MAR-concentration}
\end{figure}

We expect that relaxed, low MAR clusters will have halos with a higher DM concentration than unrelaxed, high MAR clusters \citep{Wechsler02,Ludlow2016}. This is because the larger the recent mass accretion rate, the more mass is deposited at the cluster outskirts, and the smaller the concentration. Thus, we similarly expect that gas concentration, and hence the concentration of the SZ signal, to be negatively correlated with the mass accretion rate. In the left panel of Fig.~\ref{fig:MAR-concentration}, we found that the SZ image concentration decreases as mass accretion rate increases as expected with $r=-0.21$. The simulated clusters show a similar correlation with mass accretion rate in the right panel.

\begin{figure}
	\includegraphics[width=\columnwidth]{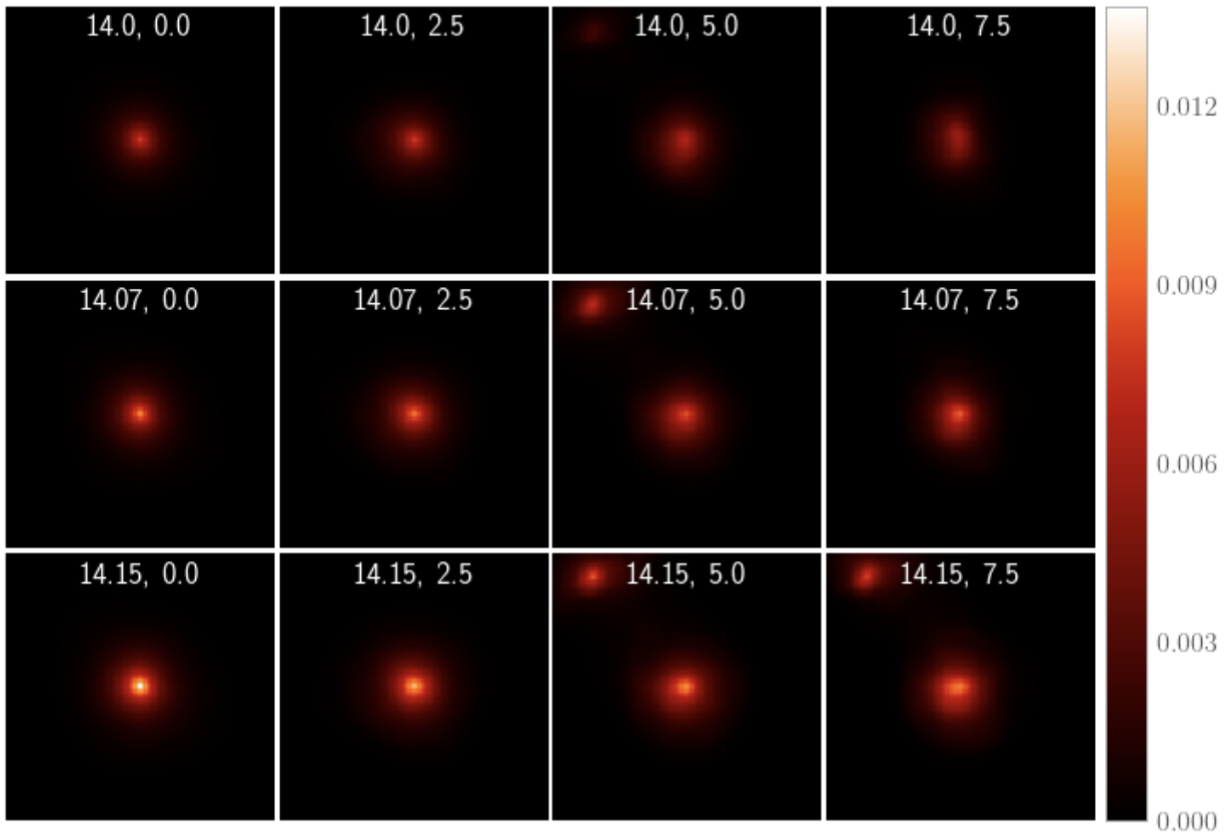}
    \caption{Examples of clusters generated by the model with increasing MAR moving from left to right and increasing mass moving from top to bottom. Each image is annotated with its mass followed by its MAR. The Gaussian noise used to create the clusters is the same for all images. All clusters are shown on a linear scale as described in Sec.~\ref{sec:data} in order to make clear the emergence of secondary halos.}
    \label{fig:MAR}
\end{figure}

In the visual appearance of clusters generated by the model we find concordant evidence for the sophisticated modeling of the effects of MAR on cluster morphology. By holding other variables constant, we can examine the effect that \textit{only} varying MAR, for example, has on the appearance of clusters. This is shown in Fig.~\ref{fig:MAR}, where it is clear that increasing MAR results in decreasing concentration and even the presence of clearly merging subhalos. Once again, this result is consistent with the physics associated with a given MAR.

\begin{figure}
	\includegraphics[width=\columnwidth]{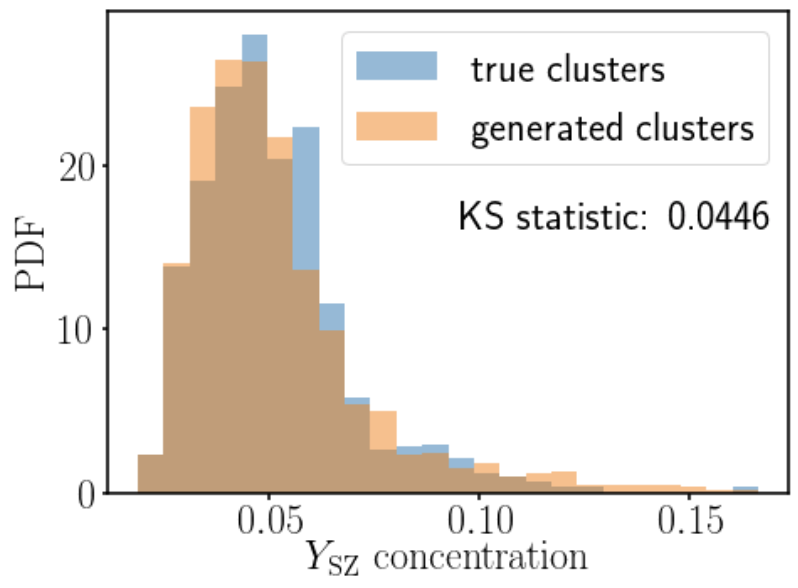}
    \caption{A comparison of the Compton $y$ parameter concentrations of the two cluster populations. We find that the model produces clusters which share a very similar concentration distribution with the simulated clusters, suggesting morphology is well-replicated.}
    \label{fig:concentration}
\end{figure}

\begin{figure}
	\includegraphics[width=\columnwidth]{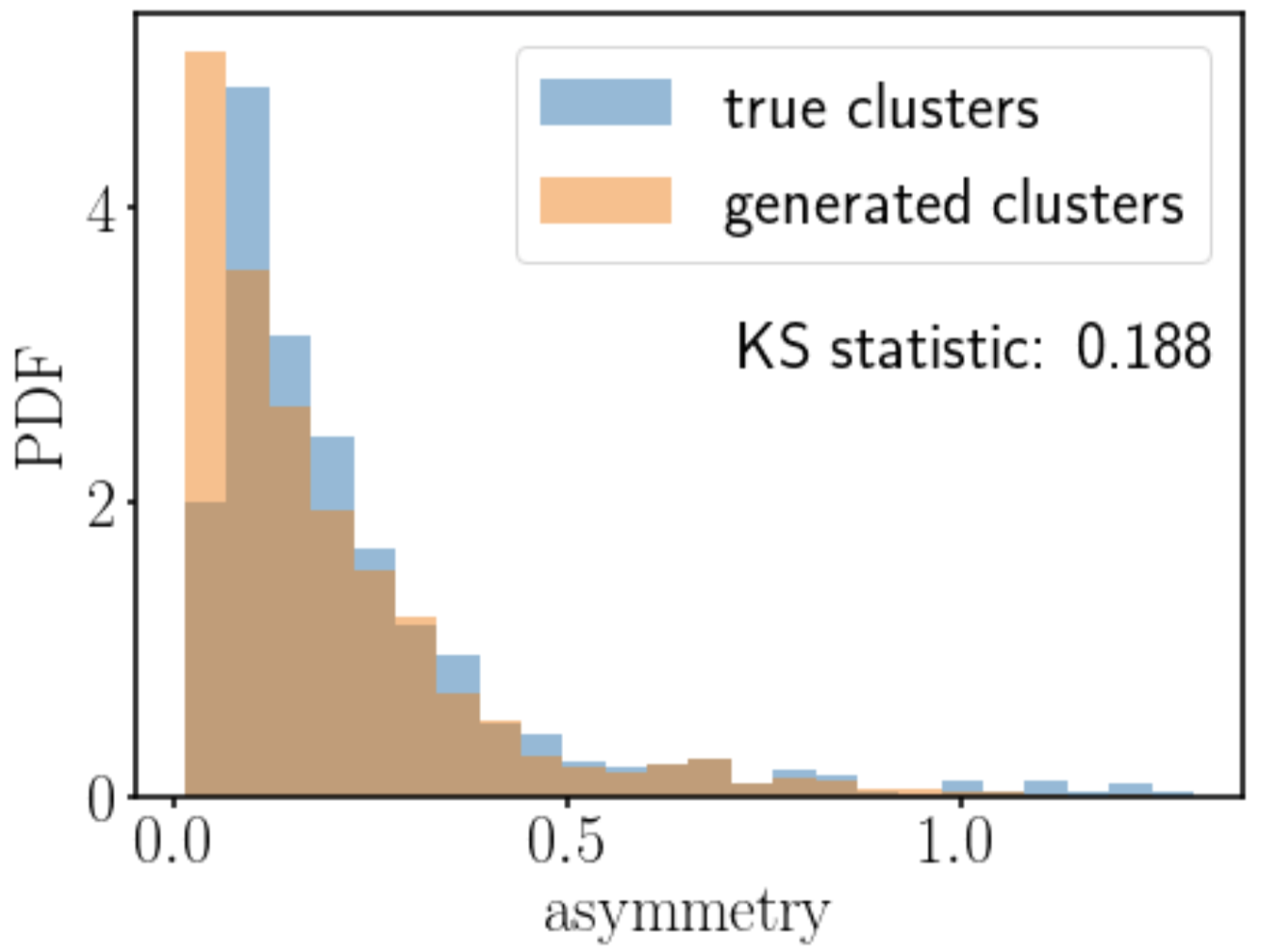}
    \caption{A comparison of the Compton $y$ parameter asymmetry of the two cluster populations. We find that the model produces clusters which share a similar asymmetry distribution with the simulated clusters, although there is a slight skewing of the generated clusters towards more symmetry.}
    \label{fig:asymmetry}
\end{figure}

\begin{figure}
	\includegraphics[width=\columnwidth]{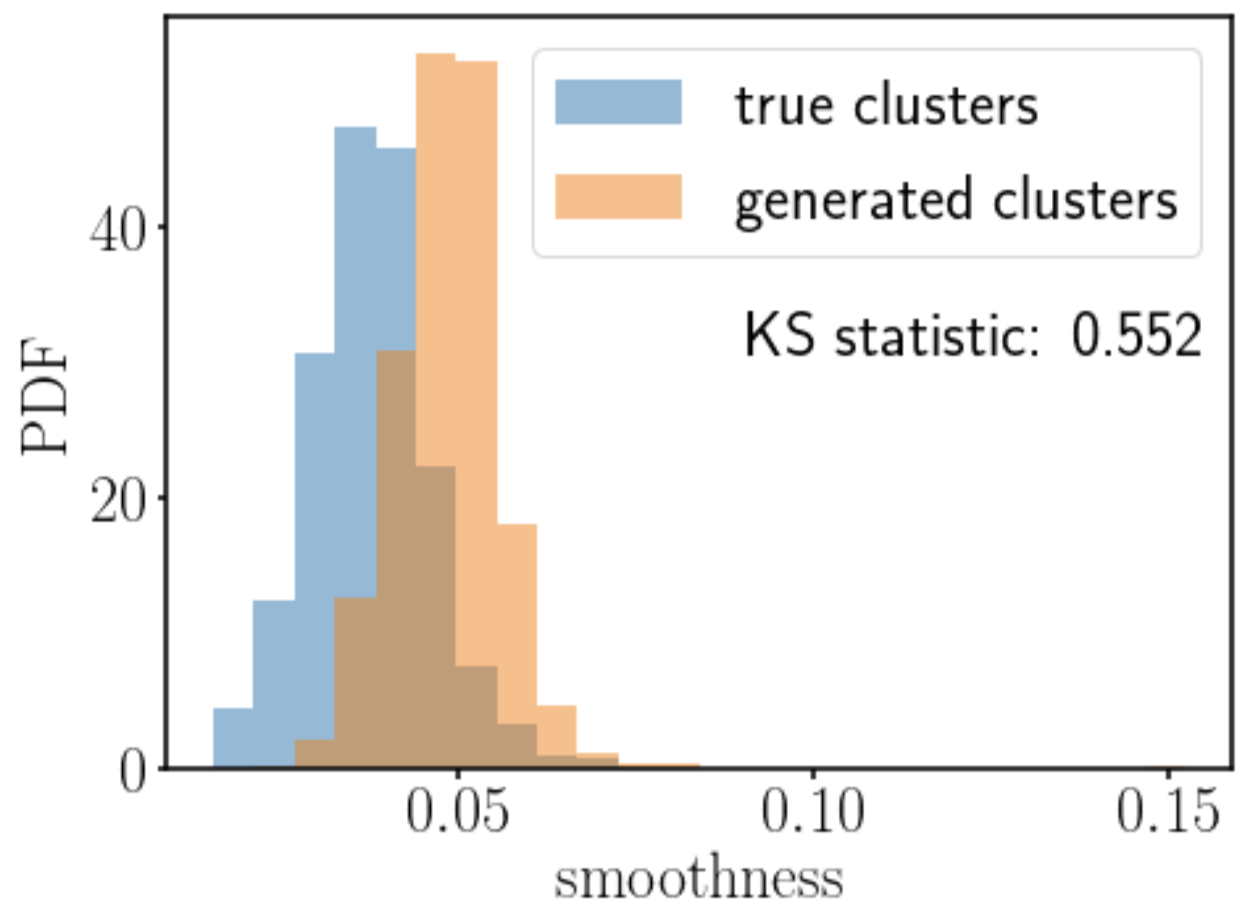}
    \caption{A comparison of the smoothness of the Compton $y$ parameter maps for the two cluster populations. We find that the model produces clusters which share a similar smoothness distribution with the simulated clusters, except that there is a prominent increase in the mean smoothness of the generated clusters.}
    \label{fig:smoothness}
\end{figure}

On first inspection the clusters generated by the model seem to be ``realistic,'' as can be seen in Fig.~\ref{fig:examples}.  But, as discussed in Sec.~\ref{sec:evaluation}, we would like to provide more substantial, ideally quantitative, support for this claim. One sense in which the generated clusters should be realistic is that they should have similar morphologies, as a population, to the morphologies of the simulated clusters. For this reason, in  Fig.~\ref{fig:concentration}, Fig.~\ref{fig:asymmetry}, and Fig.~\ref{fig:smoothness} we compare the morphological characteristics (concentration, asymmetry, and smoothness) of the two populations of clusters using histograms created with 25 equally sized bins with edges spanning from the minimum to the maximum parameter value in each case. We qualitatively assess the similarities of the distributions, but we also employ the KS statistic to quantitatively compare the similarities, providing a benchmark for future work.

As previously discussed, these distributions are expected to be conditional on the distribution of MARs and masses, so when comparing morphological distributions we use a novel cluster population with the same MAR and mass distribution as that of the simulated clusters. We find that for all morphological parameters, the general shape of the distribution is reproduced. For example in Fig.~\ref{fig:asymmetry} we find that the asymmetry distribution of the both cluster populations are strongly right-skewed with a long tail of high asymmetry. Any consistent morphological differences between the simulated and generated clusters only appeared when analyzing the asymmetry in Fig.~\ref{fig:asymmetry} (KS statistic  $\approx 0.19$) and the smoothness in Fig.~\ref{fig:smoothness} (KS statistic  $\approx 0.55$), unlike the concentration in Fig.~\ref{fig:concentration} (KS statistic  $\approx 0.045$) for which there was no discernible difference. We hypothesize that the slight over-skewing of the generated clusters towards symmetry in Fig.~\ref{fig:asymmetry} is due to the fact that the CVAE's compression necessitates some information loss, and symmetrical clusters contain much less information. The cause of the higher smoothness of generated clusters shown in Fig.~\ref{fig:smoothness} is similar to the reason why the clusters tend to be slightly more symmetrical: the compressed representations of clusters created by the CVAE are unable to capture the finest details of the cluster which are the origin of the lower smoothness. This is a well-documented tendency of VAEs: they produce blurred reconstructions and thus generate somewhat blurred images \citep{perceptualSimilarity}. Whether or not this is significant depends on the use case of the model. In this work, the particular use case we have in mind is generating predicted SZ maps for cosmological inference so we must also account for observational effects. Adopting a basic assumption of a Gaussian PSF from a CMB-S4-level instrument with $1$ arcmin resolution \citep{cmbS4} observing a clusters at a redshift of $z = 0.5$, we find that the net effect of the model's smoothing is lower than that produced by this theoretical instrumentation. Additionally, we find that the current level of smoothing is an improvement over semi-analytic models and is sufficient for the goals of this work. Higher resolution maps might be required for far-future instrumentation or different use cases (detailed analysis of substructure in a smaller number of low redshift systems, for example), and so decreasing the smoothness introduced by our model could still be an avenue of future work. 

On the whole, our morphological analysis provides further evidence that the model is in fact creating \textit{realistic} SZ maps of novel galaxy clusters.

\begin{figure*}
	\includegraphics[width=0.98\textwidth]{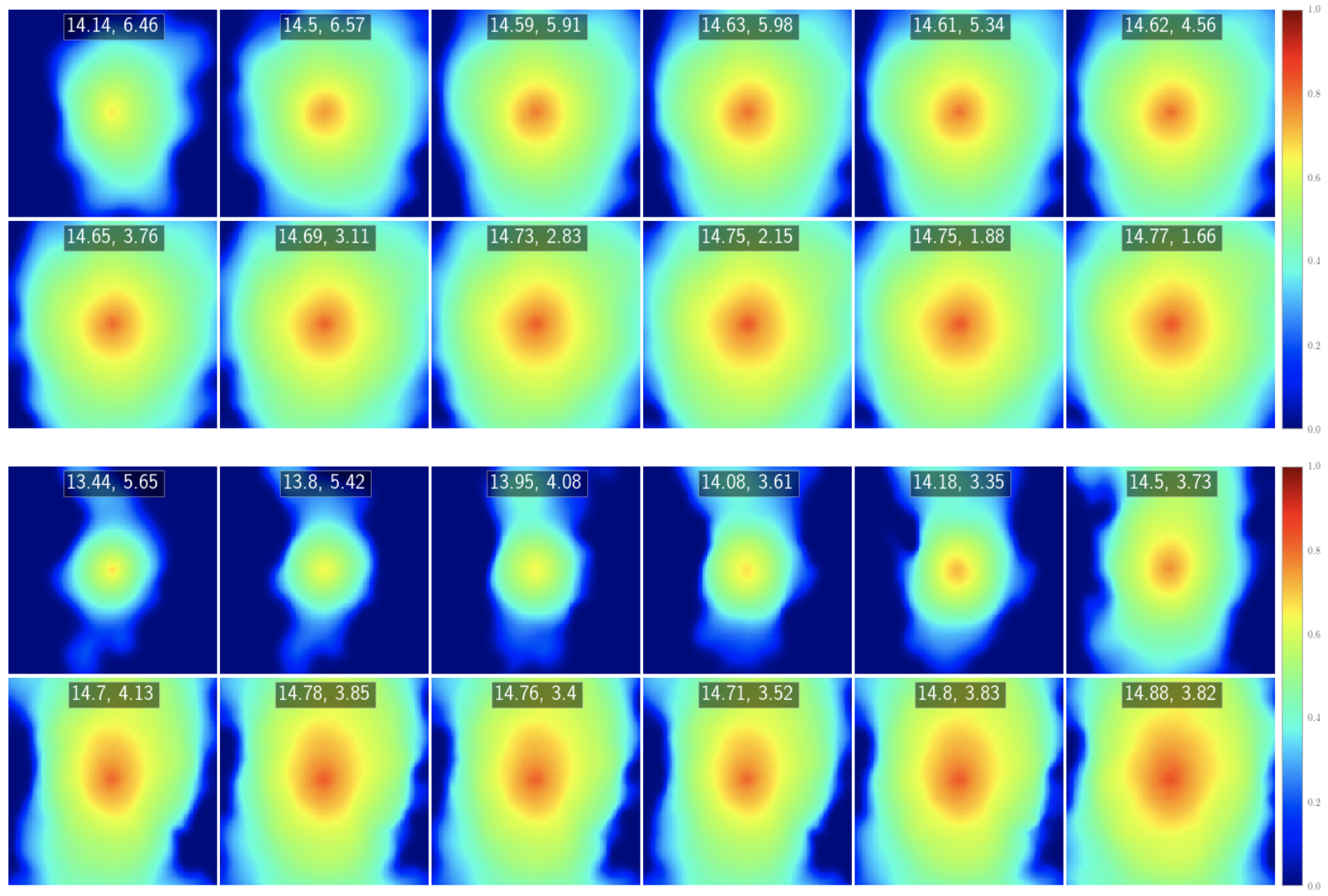}
    \caption{\textit{Top two rows}: fictional 'evolution' of a single cluster's SZ map given a relatively smooth accretion. \textit{Bottom two rows}: the same as above, except with a relatively violent accretion history. The redshifts of the maps in both sets of two rows, left to right and top to bottom, are: 1.50, 1.21, 0.997, 0.817, 0.676, 0.546, 0.440, 0.348, 0.261, 0.180, 0.110, 0.049.}
    \label{fig:evolution}
\end{figure*}

\subsection{Cluster evolution}
Our model is in some ways an explicitly atemporal model, in that all of the training data is from $z=0$. However, there is sufficient training data to show the model clusters in a wide range of dynamical states. Given this, we hypothesized that there could be some value in examining the extent to which the model could be used for simulating the evolution of a single cluster through time. To this end, we sampled the mass and MAR histories of two clusters from the training data i.e., from the Illustris-TNG simulation. The two clusters were specifically chosen to provide illustrative examples of a cluster with a smooth accretion history and a cluster with a relatively violent accretion history. These mass and MAR histories, along with a fixed latent space representation, were fed into the model to create two sequences of SZ maps corresponding to each cluster. Note that as redshift is not a parameter to the model, the apparent time dependence arises from the temporally ordered sequences of masses and MARs. We present SZ maps from between $z \approx 1.5$ and $z \approx 0.05$ for the two clusters in Fig.~\ref{fig:evolution}. From the figure, it is clear that on a qualitative basis, the model is able to realistically render the evolution of clusters through time. We mean this in the following sense: the clusters' visible growth and morphology are consistent with their mass and MAR history - see for example the evolution of the first cluster in Fig.~\ref{fig:evolution}, where we see a consistent and generally smooth increase of mass over time, consistent with given accretion history. This result leads us to come to at least two significant conclusions. Firstly, it once again indicates that the model has learned to encode some cluster physics in a physically meaningful way. If this were not the case, there would be no reason to suspect that the process described above would produce a sequence of images which appear as if they describe the consistent evolution of a \textit{single} cluster through time. Secondly, this result suggests further work is possible in learning to simulate the evolution of cluster observables.

\section{Discussion}
\label{sec:discussion}

\concsection{Physical interpretability}
Our findings in the previous section suggest that the model is able to replicate many salient properties of clusters from simulation. For context, the primary way thus far to generate SZ profiles given only a halo mass is with analytic profiles as in \cite{szGenNFW}. Recently, in \cite{greenMARSZ}, it was shown that if the MAR was also incorporated into a semi-analytic model, $70\%$ of the scatter in the $Y_{SZ}-M$ relation was modeled. By taking an ML-based approach, we have been able to replicate virtually all of the scatter, as well as morphological characteristics and even merging subhalos. From Fig.~\ref{fig:MAR-residual} we can see that some of the replication of scatter arises from accurate modeling of MAR but it is likely that some scatter is also modelled more implicitly i.e., in a way that is not as readily physically explained. Note that this stands in contrast with how existing semi-analytic models work \citep{ks01SA, greenMARSZ}. The expectation is that modeling of these extra features should improve the strength of our theoretical predictions. 

But ML-based models can do more than simply match our pre-existing expectations of physical effects, because they require very few physical assumptions. For example, we did not give the model an explicit physical description of MAR and its effects, and yet it learned to utilize MAR appropriately. In the training data, we cannot find clusters with different MARs but precisely the same mass, making it somewhat difficult to manually isolate and identify the effects of MAR. But the model's ability to interpolate and extrapolate allows us to use it to create visualizations like Fig.~\ref{fig:MAR} which clearly reveals the effects of MAR. So if we did not previously understand the relationship between MAR and cluster morphology, we could begin to understand it using the trained model. By conditioning another version of our model on cluster parameters which are more poorly understood, such as AGN feedback, or on cosmological parameters such as $\Lambda$, one could try to learn astrophysics or cosmology directly from a trained model. 

\concsection{Computational efficiency}
It should be acknowledged that the model presented in this paper has in many ways a more limited scope and focused purpose than cosmological simulations, and thus it should be expected that it would require fewer resources to run. Concretely, the TNG300-1 hydrodynamical simulation from which our training data derives required 34.9 million CPU hours to create \citep{illustrisTNG2}, not to mention the time required to create SZ maps from the simulation data. In contrast, the ML model presented in this work generates an order of magnitude more clusters in 30 seconds.

Even when compared to recent state-of-the-art ML-based methods for transforming dark matter maps into SZ maps \citep{paintingBaryons, teachingNeuralNetworks}, the model introduced in this work requires no computationally-intensive simulation data once trained. Concretely, the previously referenced ML-based models require simulated matter density maps to function (for example, from N-body simulations), while the one introduced here simply needs two numbers: mass and MAR. As discussed in Sec.~\ref{sec:intro}, while this model can be run on the outputs of N-body simulations like existing ML-based approaches, one could also achieve significant cosmological constraints by using a halo mass function to determine the cluster counts and masses to use as input to our model in a halo-based approach \citep{multiWavelengthShirasaki}.

\concsection{Stochasticity}
Guarantees on the outputs of our model, and of deep-generative models in general, are often very elusive (see \cite{deepMindGen} for related investigative work). This means that it is difficult to be certain that outputs of the model will always look as reasonable as presented here. We sample a large number of clusters from the model to ameliorate this concern, but because of the stochasticity and complexity of the model, it is still possible for it to produce a cluster of a new, non-physical form that we have not yet seen. This is simply not an issue for analytical models, which have precise descriptions of what kinds of outputs they can produce. But given that the success in reproducing various cluster features (e.g., concentration) suggests an absence of a net bias, we think it is unlikely that a few potential outliers should significantly affect the predictive utility of the model. This is still, however, a reason that ML-based methods should be approached with caution. On the other hand, because two clusters with the same mass and MAR can still have quite different morphologies, the stochasticity of our approach is also a strength. This idea is not currently captured in semi-analytical models or even some ML-based approaches \citep{teachingNeuralNetworks} because they lack stochasticity, but it is fully incorporated into our work. Concretely, given the same mass and MAR, our model can produce two clusters which are quite distinct. This behavior is more consistent with reality than previous deterministic approaches.

\begin{figure}
	\includegraphics[width=\columnwidth]{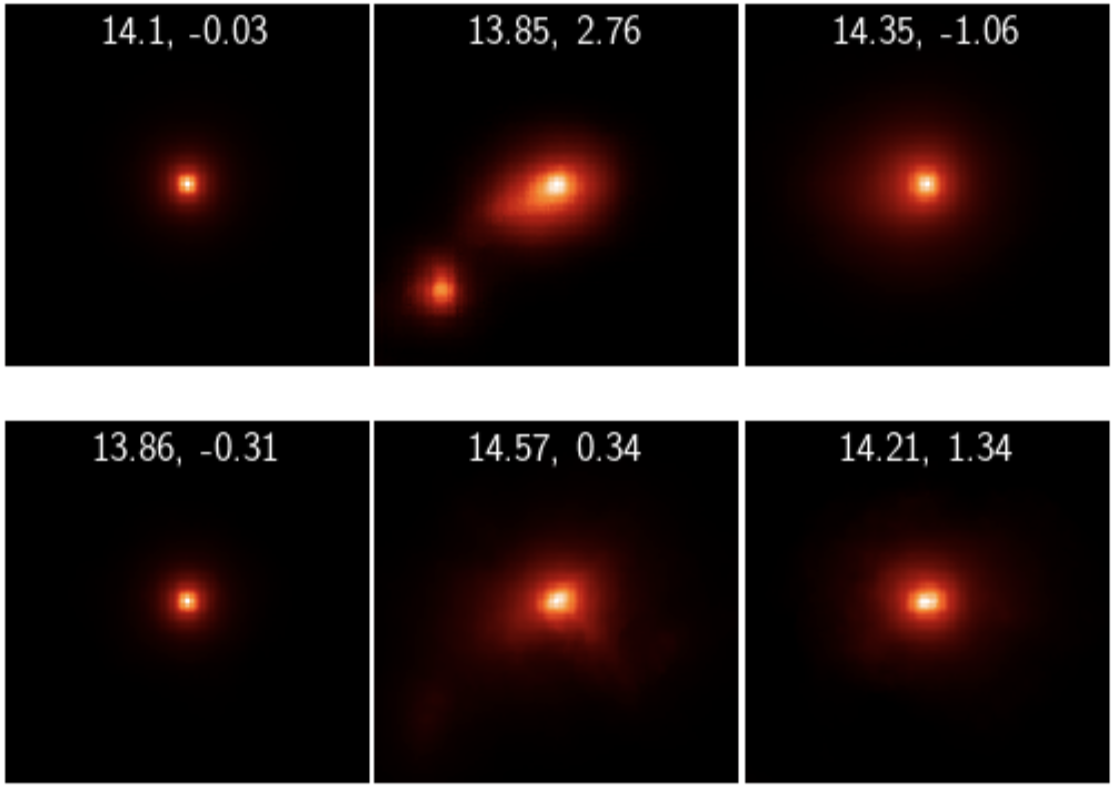}
    \caption{The top row shows randomly selected clusters which were generated by the model. The bottom row shows the "closest" cluster in the training set to the generated cluster directly above. "Closest" refers to the cluster in the training set with the smallest Euclidean pixel-wise distance from the generated cluster. Every image is normalized by dividing by its maximum pixel value before computing the Euclidean distances.
    For reference, above each cluster we list the mass followed by the MAR.}
    \label{fig:novelty}
\end{figure}

\concsection{Over-fitting}
Another potential pitfall that we explicitly sought to avoid was over-fitting, as outlined in Sec.~\ref{sec:intro}. Over-fitting occurs when the clusters generated by the model are too similar to individual examples from the training set. Another way we refer to avoiding over-fitting is to say we wish for the clusters to be ``novel'', and our methodology for assessing our success in meeting this goal is outlined in Sec.~\ref{sec:evaluation}. This is of particular concern in work like ours that uses a smaller training set than those often used in the wider field of ML. Our results are presented in Fig.~\ref{fig:novelty}. What we find is that the top row and bottom row are distinct, particularly for this unrelaxed cluster with an asymmetrical morphology. Identical rows would indicate the model is simply memorizing the training set instead of creating new (but realistic) SZ maps as we would hope. Therefore the fact that the top and bottom rows are non-identical means the model is unlikely to be over-fitting. In other words, the generated clusters do appear to be sufficiently novel.

While we perform checks to attempt to ensure the novelty of the generated cluster images, \cite{evalGenModels} points out that this technique of examining nearest neighbors is imperfect because 1) we are limited by how many examples we can inspect and 2) it is possible that a generated image is nearly a duplicate of an image in the training set, but is still far away in pixel space. Fortunately, for the use case of generating mock SZ maps of the sky, the seeming lack of bias in our results suggests that this model will be improvement over existing solutions, even if there exists a tendency to produce clusters which are more particular to our specific dataset i.e., the IllustrisTNG 300-1 simulation.

\concsection{kSZ effect} 
This work neglects to model the kSZ effect which must be eventually either included in the SZ maps or removed from observations when analyzing tSZ signals. 
Furthermore, the internal velocities in the ICM could be comparable to the overall cluster peculiar velocity for merging clusters and 20\%-30\% of the sound speed even when a cluster is relatively relaxed \citep{Nagai2003kSZ}. 
The high-resolution capability of our technique means that it is possible to obtain additional information about cluster dynamics from  kSZ substructures observed using high-resolution SZ observations \citep[e.g.,][]{Sayers_2019}. Since our method should be just as skillful at modeling the kSZ effect as modeling the tSZ effect, it would be interesting to retrain the model on the high-resolution kSZ imaging data with adjusted hyperparameters. Additionally, we expect that the simultaneous reproduction of tSZ and kSZ effects for a single halo is readily achievable using multi-channel input images. This is analagous to the common practice of training CNNs on color images in the visible spectrum.

\concsection{Observational effects}
Our work has focused on modeling the tSZ signal in the FoF region, capturing the gas that is directly associated with the primary halo. However, the diffuse gas lying outside of the FoF region (such as other halos, filaments, and sheets in the cosmic web) can contribute non-negligible tSZ signals \citep{Bregman07,Hojjati15}. We caution that these projection effects must be accounted for using the full-lightcone simulations before applying our algorithm to observational data analyses, especially in low-mass clusters and groups that are significantly affected by the projection effects \citep{Hallman07}.  

\concsection{Alternative algorithms}
As outlined previously, we use a machine-learning--based technique in this work because of the efficiency and scalability of a trained model. For example, a working model can very quickly generate mock clusters for an entire light cone using a single processor. However, given the choice of a machine learning approach, there are still several possible architectures for generative modeling. The data's natural representation as images suggested convolutional neural networks as an obvious choice. There are then two main alternatives: the conditional generative adversarial network (CGAN) \citep{ganOriginal} and the CVAE. On a practical level, CGANs are often more difficult to train because of optimization instabilities associated with training two separate networks (i.e., mode collapse), although there are have also been a variety of attempts to remedy these difficulties \citep{improvingGANs}. CVAEs also possess the nice property that they explicitly create encodings of the training dataset, which can then be used for a variety of separate inference or general exploratory tasks. Ultimately these two facts led us to use a CVAE in this work, however it is reasonable to expect that a CGAN could perform a similar task.

\concsection{Limitations of simulation-based approaches}
A primary concern of simulation-based approaches such as ours is that any biases in the simulations - perhaps due to poorly modeled physical processes - could find their way into our ultimate predictions. But several aspects of our particular use case ameliorates this concern. First, the impacts of baryonic effects on massive cluster-size halos are much better understood and controlled for than the effects on galaxies and groups. Secondly, the thermal SZ effect signal gets little contribution from cluster cores that are affected significantly by the poorly understood feedback physics. In fact, there have been significant recent developments in our understanding of the physics of galaxy cluster outskirts, relevant for modeling SZ surveys, from both simulations and observations (see \cite{walkerclusterphysics} for a recent review). Building on these new insights, we focused on modeling the major sources of systematic uncertainties, such as non-thermal pressure and gas shapes, that are known to make a significant impact on the thermal SZ signal even for massive clusters. Despite these recent improvements, it would still be important to test this technique using the plethora of hydrodynamical simulations of galaxy cluster formation (such as Magneticum, Bahama/MACSIS, and TheThreeHundred) that have recently become available. For eventual cosmological applications, it would be especially important to perform and analyze a suite of zoom-in hydrodynamical cosmological simulations of galaxy clusters for a broad range of astrophysical and cosmological parameter space using multiple subgrid models of galaxy formation, analogous to CAMELS \citep{camels,camels-mfd} on galaxy formation, which provide a much needed training set for cluster cosmology.

\section{Conclusions}
\label{sec:conclusions}
In this work, we train a deep learning model to generate realistic high resolution SZ maps of galaxy clusters given a mass and mass accretion rate (MAR). We find that in many respects the model succeeds in meeting our goals. Specifically:

\begin{itemize}
\item We see that the the model is able to create clusters whose $Y-M$ relation has identical slope and scatter (with differences of $<1\%$) to the $Y-M$ relation of simulated clusters (Fig.~\ref{fig:Y-M_norm}, Fig.~\ref{fig:Y-M_scatter}). This indicates that the model explicitly takes into account baryonic physics, structure formation processes and projection which can change the slope, normalization and scatter of $Y-M$ relations. Meanwhile, semi-analytic models, which only take into account mass accretion processes, can only reproduce $30-50\%$ of the scatter found in simulations.

\item Our analysis suggests that the model correctly models the effects of the mass accretion rate on SZ images. We demonstrate that the model produces clusters which exhibit a correct correlation between MAR and the residuals of the best fit $Y-M$ relation (Fig.~\ref{fig:MAR-residual}) or the concentration of SZ images (Fig.~\ref{fig:MAR-concentration}). Visually, we can see that given high mass accretion rate, the model will generate a more unrelaxed, even merging, clusters (Fig.~\ref{fig:MAR}). All show that the expected physical consequences of MAR are reflected in the clusters generated by the model.

\item We find by a direct examination that the morphologies of clusters from a sophisticated magnetohydrodynamical simulation are well matched by the morphologies of our novel clusters. The Compton $y$ parameter spatial concentration of clusters is clearly well emulated by the model (Fig.~\ref{fig:concentration}). The model demonstrates similar success is recreating the asymmetry of clusters (Fig.~\ref{fig:asymmetry}). In particular, note that the model makes no simplifying assumptions about symmetry when generating clusters and is inherently stochastic. We also find that the smoothness of the maps generated by the model suggests a sufficient resolution for use with upcoming SZ survey results (Fig.~\ref{fig:smoothness}).
\item The mass and MAR dependence of the SZ maps are captured with sufficient sophistication to create convincing imitations of cluster evolution through time, even in the absence of training data with a temporal component (Fig.~\ref{fig:evolution}).
\item When comparing the clusters generated by the model with training clusters from simulation, we find an absence of direct duplication (Fig.~\ref{fig:novelty}). This suggests the model is creating somewhat novel clusters and is not simply duplicating what it has seen before.
\end{itemize}

We have shown that machine-learning--based methods are a viable option for generating large amounts of high resolution SZ images of galaxy clusters, where the features due to sophisticated physics are explicitly reproduced. We envision a model like this one primarily being used in a halo-based approach to the generation of observables. With this approach, the ability to cheaply generate hundreds of thousands of realistic SZ images of galaxy clusters in seconds for halo catalogs corresponding to different cosmological models should enable the extraction of tighter cosmological constraints and information about potentially new physics from upcoming microwave surveys.

\section*{Acknowledgements}
We gratefully acknowledge the \textsc{IllustrisTNG} Team for publicly releasing all TNG simulation data \citep{illustrisTNG-DR}. The authors would like to thank the open source community for their work on the Keras library (\url{http://www.keras.io}) which was used in this work. This work was supported in part by the facilities and staff of the Yale Center for Research Computing. SBG is supported by the US National Science Foundation Graduate Research Fellowship under Grant No. DGE-1752134.

\section*{Data Availability}
The Illustris TNG simulation data is publicly available here: \url{https://www.tng-project.org/data/}. SZpack, used to generate the corresponding SZ maps used in this work, can be found here: \url{http://www.jb.man.ac.uk/~jchluba/Science/SZpack/SZpack.html}. Additionally, the code written by the authors for this work is available upon request. 



\bibliographystyle{mnras}
\bibliography{biblio} 


\bsp	
\label{lastpage}
\end{document}